\documentclass{aa}  
\usepackage{graphicx}
\usepackage{caption}
\usepackage{subcaption}
\usepackage{txfonts}

\newcommand{\wx}{\omega_x}
\newcommand{\wy}{\omega_y}
\newcommand{\wz}{\omega_z}

\newcommand{\wxp}{\omega_{x'}}
\newcommand{\wyp}{\omega_{y'}}
\newcommand{\wzp}{\omega_{z'}}

\newcommand{\wps}{\omega_{p,\circlearrowright}}
\newcommand{\ips}{i_{p,\circlearrowright}}

\newcommand{\pdd}[2]{\frac{\partial #1}{\partial #2}}

\usepackage{hyperref}
\usepackage{siunitx}
\providecommand{\orcit}[1]{\protect\href{https://orcid.org/#1}{\protect\includegraphics[width=8pt]{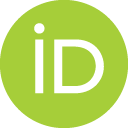}}}

\begin{document} 

   \title{Rotational broadening of exoplanet spectra in arbitrarily oriented systems: Application to reflected light}

   \titlerunning{Rotational broadening in arbitrarily orientated systems}

   \subtitle{}

   \author{Sophia R. Vaughan\orcit{0000-0002-8199-9818}\inst{1} and Laura Kreidberg\orcit{0000-0003-0514-1147}\inst{1}}

   \institute{Max-Planck-Institut f\"ur Astronomie, K\"onigstuhl 17, 69117 Heidelberg, Germany
             }

   \date{Received 29 June 2026; accepted 25 August 2026}
 
  \abstract
   {Exoplanet reflection spectra contain a wealth of information, which we are now beginning to access. At a high spectral resolution, the rotation of both the star and the planet affects the shape of the reflected spectral lines. This effect must be accounted for in high-resolution analyses and could be used to measure the orbital alignment of systems. In this work, we derive the rotational broadening kernels for arbitrarily aligned systems, both for the stellar spectrum viewed by the planet and for the reflected component of the planet's spectrum. We then determine the relation between the parameters of the kernels and the standard orbital parameters of an exoplanet system. Finally, we show examples of these kernels, compare them with previous models, and investigate the limitations imposed by our assumptions. We find that the difference in the rotational broadening of a non-aligned system relative to the broadening expected from a prograde aligned orbit can be of the order of $\SI{50}{\kilo\meter\per\second}$. We also highlight how the equations and kernels presented in this work could be modified for other uses, such as thermal emission spectra.}
   {}{}{}{}


   \maketitle
%

\section{Introduction}

Exoplanet characterisation is a rapidly growing field of research which uses a variety of techniques to extract the transmission, emission, and reflection spectra of exoplanets and derive information on their atmospheres. Measurements of reflection spectra are less common because the planets currently most amenable to characterisation, the hot-Jupiters, typically have low albedos \citep[$A_g\sim0.1$; e.g.][]{Esteves2015}. However, there are exceptions, for example, Kepler 7\,b , HD189733\,b, and LTT-9779\,b \citep{Demory2011, Evens2013, Hoyer2023}. Near-future facilities such as the Nancy Grace Roman Space Telescope, the PLAnetary Transits and Oscillations of stars telescope (PLATO), and the Extremely Large Telescope (ELT) will soon enable the characterisation of a larger number of exoplanets in reflected light\citep[e.g.][]{Rauer2014, Carrion-Gonzalez2021, Vaughan2024, Palle2025}. 

At high spectral resolution, the spectra become sensitive to the rotational properties of the system through rotational broadening of the spectral lines. This is an important effect to account for when obtaining a high-resolution spectrum of an exoplanet because current techniques typically use cross-correlations with model templates \citep[][and references therein]{Birkby2018, Snellen2025}. The efficiency of these techniques is highly dependent on the broadening of the spectral lines \citep[e.g.][]{Spring2022}.

The rotational broadening kernels used to broaden reflected stellar spectra are based on the rotational broadening kernels for stars \citep[e.g.][]{Carroll1928, Shajn1929, Gray2022}. Previous studies have adapted these to model the rotational broadening kernel for a planet at full phase on a prograde aligned orbit \citep[e.g.][]{Rodler2010, Spring2022} and have approximated other scenarios \citep{Winterhalder2026}. Here we derive the full analytical rotational broadening kernel for the reflection spectrum of an exoplanet on an arbitrarily orientated orbit. In Section \ref{sec:2} we discuss the origins of the rotational broadening of the reflected light. We derive the kernels in Section \ref{sec:kernels} and apply them to the orbital parameters of an exoplanet system in Section \ref{sec:toparmaters}. In Section \ref{sec:discussion}, we show examples of the kernels, compare them to previous approximations, and discuss the impact of our assumptions. We conclude in Section \ref{sec:conclusions}. 

\section{Rotational broadening in reflection spectra}
\label{sec:2}

The reflection spectrum of a planet includes spectral lines from two sources: the light source spectrum and the planet's albedo modulation. The reflected stellar spectrum is not the same as the spectrum of the star measured from Earth. This is because, due to the planet's orbital motion, the star appears to rotate at a different rate from the perspective of the planet, and thus the reflected spectrum has a different rotational broadening. We define $f_\star(\lambda)$ as the unbroadened stellar spectrum and $k_\star(v)$ as the stellar rotational broadening kernel constructed based on the stellar rotation viewed from the planet. The stellar rotational broadening kernel can be considered as a smoothing function for the stellar spectrum. It acts on the stellar spectrum via a convolution, and therefore the Doppler velocity, $v$, needs to be converted to the corresponding wavelength shift, $\Delta\lambda$. This can be done assuming the non-relativistic formula $v = c \frac{\Delta\lambda}{\lambda}$. Typically, $\lambda$ is assumed to be fixed to the central wavelength of the spectrum, which means that the kernel is constant as a function of wavelength. However, caution is required for spectra that span large wavelength ranges. Once converted, the stellar spectrum viewed by the planet can be written as the convolution of the stellar spectrum and the broadening kernel, $ f_\star(\lambda) * k_\star(\Delta\lambda)$.

Exoplanets do not reflect light equally at all wavelengths. The amount of light that a planet reflects as a function of wavelength is encoded in its albedo, $A_g(\lambda)$. Therefore, the reflected spectrum is the stellar spectrum viewed by the planet, weighted by the albedo, $A_g(\lambda) ( \; f_\star(\lambda) * k_\star(\Delta\lambda) \; )$. However, this is not the spectrum that we observe from Earth. The exoplanet spins on its axis, which rotationally broadens the reflected spectrum, but the broadening is affected by the portion of the planet's visible disc that is illuminated. We define the rotational broadening resulting from the planet's own spin using the kernel $k_p(v)$. We refer to this as the planet's broadening, although we note that it affects the stellar spectral lines as well. Again, this kernel must be converted from the Doppler velocity, $v$, to the corresponding wavelength shift, $\Delta\lambda$, before the convolution. The spectrum we observe from the exoplanet, $f_p(\lambda)$, is given by the convolution of the planet's rotational broadening kernel with the spectrum reflected by the planet, $f_p(\lambda) = ( \; A_g(\lambda) ( \; f_\star(\lambda) * k_\star(\Delta\lambda) \; ) \; ) * k_p(\Delta\lambda)$.

\section{Computing the rotational broadening kernel}
\label{sec:kernels}

\subsection{Defining the rotational broadening kernel}

To compute the rotational broadening kernel for the visible hemisphere of a spherical object, we evaluated the following integral over the surface of the hemisphere $A$:

\begin{equation}
    \label{eq:rotational_kernel}
    k(v) = \iint\limits_{A} I(x,y,z) \delta(v - v_{los}(x,y,z)) \mathrm{d} A,
\end{equation}
where $I(x,y,z)$ is the surface brightness of the hemisphere, $v$ is the velocity variable, and $v_{los}(x,y,z)$ is the line-of-sight velocity across the surface of the sphere. Here, $\delta$ is a Dirac delta with the following form:

\begin{equation}
    \delta(\chi) = 
    \begin{cases}
        1 \quad \text{if } \chi=0 \\
        0 \quad \text{otherwise}
    \end{cases}.
\end{equation}

This function models spectral lines as infinitely narrow. However, we can still recover the kernel for finite line shapes by convolving the non-rotationally broadened line shape with this rotational kernel (see Appendix \ref{sec:proof_narrow_lines}). 

This rotational kernel is un-normalised. When rotationally broadening a spectrum, the kernel must sum to unity to conserve flux. This normalisation depends on the sampling of the kernel. Therefore, in this work, we only present only un-normalised kernels. Equation \ref{eq:rotational_kernel} may not always be analytically solvable. However, for the rest of this section, we discuss several cases relevant to the rotational broadening of the reflected light from exoplanets that are analytically solvable. 

\subsection{The kernel for spatially constant angular velocity}
\label{sec:uniform_ang}

We assumed that the motion of the spherical surface can be described by an angular velocity vector $\vec{\omega} = (\wx, \wy, \wz)$, and therefore the velocity is $\vec{v} = \vec{\omega} \times \vec{r}$. We assumed that $\wx$, $\wy$, and $\wz$ are constant with respect to the spatial coordinates $x$, $y$, and $z$. That is, the sphere rotates as a solid body. We discuss the impact of this assumption in Section \ref{sec:approx_discussion}. In this formulation, we chose to place the observer at a distant point on the positive x axis. This is convenient for relating the stellar rotational broadening kernel, for which the planet is the observer, to conventional orbital parameters, as discussed in Section \ref{sec:map_stellar}. For the planetary rotational broadening kernel, it is more convenient to relate the kernel to conventional orbital parameters when the observer, which in this case is Earth, is on the positive z axis (see Section \ref{sec:map_planet}). We present a re-derivation of these equations for an observer on the positive z axis in Appendix \ref{sec:zobserver}. 

When the observer is on a distant point on the positive x axis, the line-of-sight velocity is $v_{los}(x,y,z) = \wy z - \wz y$. The un-normalised rotational kernel is

\begin{equation}
    k(v) = \iint\limits_{A} I(x,y,z) \delta(v - \wy z + \wz y) \mathrm{d} A.
\end{equation}

This surface integral is over the surface of the observer-facing hemisphere defined by $x>0$. It is possible to simplify this integral into a single integration over a dummy variable $\phi$. First, we converted this integral into a volume integral by adding a second Dirac delta function that ensures that only the surface of radius $R$ contributes to the integral (see Equation \ref{eq:dirac_property}). We determined the integration limits of $x$ from the hemisphere condition defined by $x>0$:

\begin{equation}
    \label{eq:volume_integral}
    k(v) = \int\limits_{-\infty}^{\infty}\int\limits_{-\infty}^{\infty}\int\limits_{0}^{\infty} I(x,y,z) \delta(v - \wy z + \wz y) \delta(x^2+y^2+z^2 - R^2) \mathrm{d} x \mathrm{d} y \mathrm{d} z .
\end{equation}

Because of the two Dirac deltas, we know that the only regions of the volume that contribute to the integral obey $x^2+y^2+z^2=R^2$ and $v - \wy z + \wz y = 0$. Substituting the latter into the former and rearranging:

\begin{equation}
    x^2+\left(\frac{\wy z - v}{\wz}\right)^2+z^2=R^2
\end{equation}
\begin{equation}
    \implies x^2+ \left(\frac{\wy^2+\wz^2}{\wz^2}\right) \left(z - \frac{\wy v}{\wz^2+\wy^2}\right)^2  = R^2 - \frac{v^2}{\wz^2+\wy^2}.
\end{equation}
This is the equation for an ellipse. This indicates that the following coordinate transform, which maps this ellipse to a circle, makes evaluating the integral easier:

\begin{equation}
    \label{eq:coordinate_tranform}
    x = r\cos{\phi}; \quad y=y; \quad z = \frac{\wz}{(\wy^2 + \wz^2)^\frac{1}{2}} r \sin{\phi} + \frac{\wy v}{\wy^2 + \wz^2}.
\end{equation}

Next, we transformed our integral into the new coordinate system defined by ($r$, $\phi$, $y$). The volume elements in each coordinate system are related via $\mathrm{d}V = \mathrm{d}x\mathrm{d}y\mathrm{d}z=\det(J)\mathrm{d}r\mathrm{d}\phi\mathrm{d}y$, where $\det(J)$ is the determinant of the Jacobian of the coordinate transform. The integral after the coordinate transform becomes

\begin{equation}
    J = \begin{bmatrix} \pdd{x}{r} & \pdd{x}{\phi} & \pdd{x}{y} \\ \pdd{y}{r} & \pdd{y}{\phi} & \pdd{y}{y} \\ \pdd{z}{r} & \pdd{z}{\phi} & \pdd{z}{y} \end{bmatrix} =
    \begin{bmatrix} \sin\phi & r\cos\phi & 0 \\ 0 & 0 & 1 \\ \frac{\wz}{(\wy^2 + \wz^2)^\frac{1}{2}}\cos\phi & -\frac{\wz}{(\wy^2 + \wz^2)^\frac{1}{2}}r\sin\phi & 0 \end{bmatrix}
\end{equation}
\begin{equation}
    \implies \det(J) = \frac{\wz}{(\wy^2 + \wz^2)^\frac{1}{2}} r
\end{equation}

\begin{equation}
    \label{eq:volume_integral_nc}
    \begin{split}
        k(v) = \int\limits_{-\infty}^{\infty}\int\limits_{0}^{\infty}\int\limits_{-\frac{\pi}{2}}^{\frac{\pi}{2}} &\frac{\wz}{(\wy^2 + \wz^2)^\frac{1}{2}} I(r, \phi, y) \delta(v - \wy z(r,\phi) + \wz y) \\ &\delta(x(r,\phi)^2+y^2+z(r,\phi)^2 - R^2)  r \mathrm{d}\phi\mathrm{d}r\mathrm{d}y.
    \end{split}
\end{equation}

We temporarily left $x$ and $z$ as functions of $r$ and $\phi$ for clarity and updated the integration limits so that the whole $x>0$ volume is included. Integrating over $y$ enforces that $y = \frac{\wy z(r,\phi) - v}{\wz}$ through the integration properties of the first Dirac delta, so the integral becomes

\begin{equation}
    \begin{split}
    k(v) =& \int\limits_{0}^{\pi}\int\limits_{0}^{\infty} \frac{\wz}{(\wy^2 + \wz^2)^\frac{1}{2}} I\left(r, \phi, \frac{\wy z(r,\phi) - v}{\wz}\right) \\ &\delta(x(r,\phi)^2+\left(\frac{\wy z(r,\phi) - v}{\wz}\right)^2+z(r,\phi)^2 - R^2)  r \mathrm{d}r\mathrm{d}\phi
    \end{split}.
\end{equation}

Substituting $z(r, \phi)$ and $x = r\cos\phi$ into the Dirac delta, we find that the integral can be rewritten as follows:

\begin{equation}
    \begin{split}
    \left(\frac{\wy z(r,\phi) - v}{\wz}\right)^2+z(r,\phi)^2 = r^2 \sin^2{\phi} + \frac{v^2}{\wy^2 + \wz^2} 
    \end{split}
\end{equation}

\begin{equation}
    \begin{split}
    k(v) = \int\limits_{0}^{\pi}\int\limits_{0}^{\infty} & \frac{\wz}{(\wy^2 + \wz^2)^\frac{1}{2}}I\left(r, \phi, \frac{\wy z(r,\phi) - v}{\wz}\right) \\ &\delta\left(r^2 + \frac{v^2}{\wy^2+\wz^2} - R^2\right)  r \mathrm{d}r\mathrm{d}\phi.
    \end{split}
\end{equation}

Integrating with respect to $r$ is now straightforward, and the remaining Dirac delta enforces $r^2 = R^2 - \frac{v^2}{\wy^2+\wz^2}$:

\begin{equation}
    \label{eq:kernel_constants}
    k(v) =  \int\limits_{0}^{\pi} \frac{\wz}{(\wy^2 + \wz^2)^\frac{1}{2}} I(\phi)  \left( R^2 - \frac{v^2}{\wy^2+\wz^2} \right)^\frac{1}{2} \mathrm{d}\phi.
\end{equation}

The function $I(x,y,z)$ is now written only in terms of $\phi$. By applying the conditions imposed by the two Dirac deltas, namely $y = \frac{\wy z(r,\phi) - v}{\wz}$ and $r^2 = R^2 - \frac{v^2}{\wy^2+\wz^2}$, onto the original coordinate system in Equation \ref{eq:coordinate_tranform}, we define the following substitutions for $(x,y,z)$ to convert $I(x,y,z)$ to $I(\phi)$: 

\begin{equation}
    \label{eq:substitutions}
    \begin{split}
        x &= \left(R^2 - \frac{v^2}{\wy^2 + \wz^2}\right)^{\frac{1}{2}} \cos{\phi} \\
        y &= \frac{\wy}{(\wy^2+\wz^2)^\frac{1}{2}}\left(R^2 - \frac{v^2}{\wy^2 + \wz^2}\right)^{\frac{1}{2}} \sin{\phi} - \frac{\wz v}{\wy^2+\wz^2} \\
        z &= \frac{\wz}{(\wy^2+\wz^2)^\frac{1}{2}}\left(R^2 - \frac{v^2}{\wy^2 + \wz^2}\right)^{\frac{1}{2}} \sin{\phi} + \frac{\wy v}{\wy^2+\wz^2}.
    \end{split}
\end{equation}

Omitting the preceding constants in Equation \ref{eq:kernel_constants}, we find that provided $\wz \neq 0$, the un-normalised kernel is

\begin{equation}
    \label{eq:unnormalised_kernel}
    k(v) = \int\limits_{-\frac{\pi}{2}}^{\frac{\pi}{2}} I(\phi)  \left( 1 - \frac{v^2}{R^2(\wy^2+\wz^2)} \right)^\frac{1}{2} \mathrm{d}\phi.
\end{equation}

\begin{figure}
    \centering
    \includegraphics[width=0.95\linewidth, trim={0cm 0.5cm 0cm 0cm},clip]{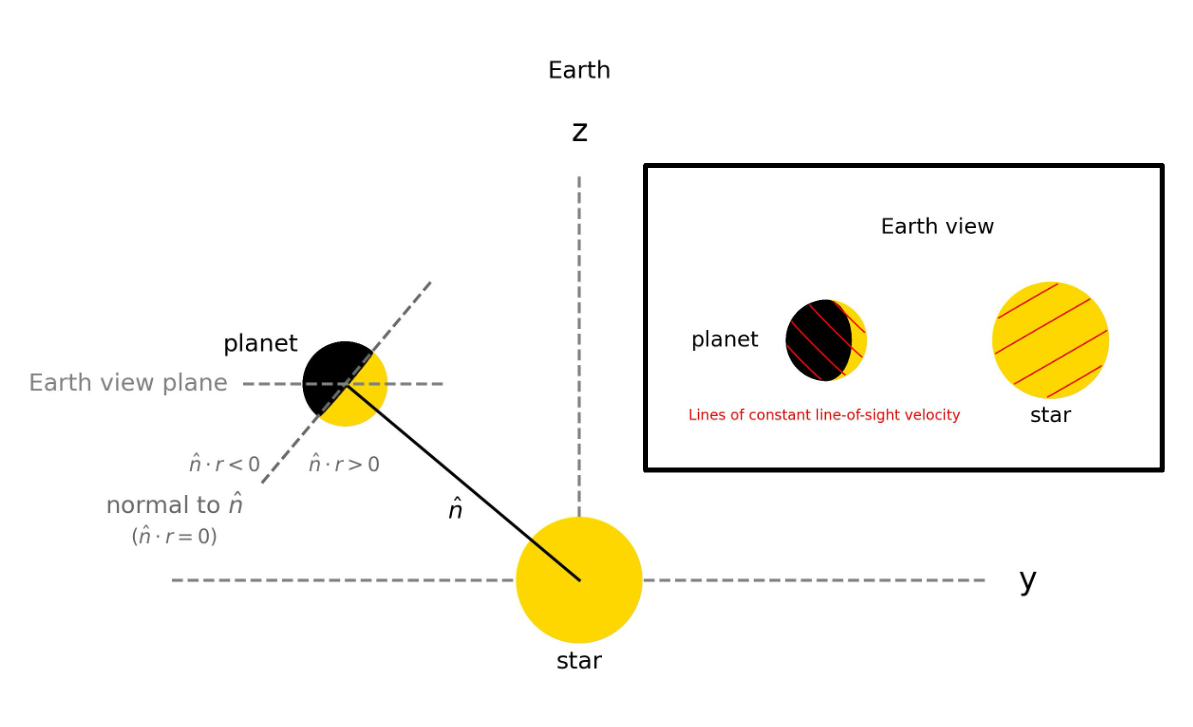}
    \caption{Illustration of a star and a planet showing how to find the day-night terminator. The inset shows the star and planet with example contours of constant line-of-sight velocity.}
    \label{fig:illustration}
\end{figure}

The variable $\phi$ is a dummy variable that defines a specific position on the surface of a sphere along a contour of constant radial velocity with respect to an observer. The red lines in the inset of Figure \ref{fig:illustration} show examples of such contours. The integral over $\phi$ measures the length of the line, and therefore by integrating the surface brightness over $\phi$, we find the fraction of light that is Doppler-shifted to a given value. Ordinarily, one would want to integrate the surface brightness throughout the visible hemisphere, that is, by integrating $\phi$ from $-\frac{\pi}{2}$ to $\frac{\pi}{2}$. 

Depending on the functional form of $I(x,y,z)$, Equation \ref{eq:unnormalised_kernel} may not be easy or possible to solve analytically. Next we investigate two simple, solvable cases that are useful when studying the reflected light of exoplanets.

\subsubsection{Assuming a uniform brightness disk}
\label{sec:uniform_disk}

If the spherical disc has uniform brightness, then $I(x,y,z) = 1$ and Equation \ref{eq:unnormalised_kernel} is trivially integrated. Therefore, the rotational kernel for a uniformly bright sphere with a spatially independent angular momentum (solid body rotation) is

\begin{equation}
    \label{eq:uniform_kernel}
        k(v) = \pi \left(1 - \frac{v^2}{R^2(\wy^2 + \wz^2)}\right)^{\frac{1}{2}}.
\end{equation}

\subsubsection{Assuming a uniform brightness crescent}
\label{sec:uniform_cresent}

We assumed that the unit vector $\vec{\hat{n}}$ points from the centre of the sphere towards a distant light source. The illuminated portion of the sphere is represented by $\vec{\hat{n}} \cdot \vec{r} \geq 0$, that is, $n_x x + n_y y + n_z z > 0$, as shown in the main panel of Figure \ref{fig:illustration}. The inset of Figure \ref{fig:illustration} shows that, when the flux is zero on the nightside, we need to integrate $\phi$ to the dayside-nightside terminator. To find the integration limits, we substituted Equation \ref{eq:substitutions} into $\vec{\hat{n}} \cdot \vec{r} = 0$ and solved for $\phi$:

\begin{equation}
\label{eq:cresent_limit_equation}
\begin{split}
    \frac{v(\wy n_z - \wz n_y)}{\wy^2 + \wz^2} + n_x \left(R^2 - \frac{v^2}{\wy^2 + \wz^2}\right)^{\frac{1}{2}} \cos(\phi) \\+ \frac{\wy n_y + \wz n_z}{(\wy^2 + \wz^2)^\frac{1}{2}} \left(R^2 - \frac{v^2}{\wy^2 + \wz^2}\right)^{\frac{1}{2}} \sin(\phi) = 0.
\end{split}
\end{equation}

For clarity, we rewrote this as $c_1 \cos(\phi) + c_2 \sin(\phi) + c_3 = 0$ and found that the equation has the following analytical solution. First, we used the double angle formulae and the identity ($\cos^2{\frac{\phi}{2}} + \sin^2{\frac{\phi}{2}} = 1$) to convert the equation to

\begin{equation}
\begin{split}
        c_1 \cos^2{\frac{\phi}{2}} - c_1 \sin^2{\frac{\phi}{2}} + 2 c_2 \sin{\frac{\phi}{2}} \cos{\frac{\phi}{2}} \\ + c_3 \cos^2{\frac{\phi}{2}} + c_3\sin^2{\frac{\phi}{2}} = 0.    
\end{split}
\end{equation}

Next, we divided the equation by $\cos^2{\frac{\phi}{2}}$. The result was a quadratic equation in terms of $\tan{\frac{\phi}{2}}$. This equation yields the following two solutions for $\phi(v)$, which depend on the velocity:

\begin{equation}
        c_1 - c_1 \tan^2{\frac{\phi}{2}} + 2 c_2 \tan{\frac{\phi}{2}} + c_3 + c_3\tan^2{\frac{\phi}{2}} = 0
\end{equation}
\begin{equation}
        (c_3 - c_1) \tan^2{\frac{\phi}{2}} + 2 c_2 \tan{\frac{\phi}{2}} + c_3 + c_1 = 0
\end{equation}
\begin{equation}
        \tan{\frac{\phi}{2}} = \frac{- c_2 \pm \sqrt{c_1^2+c_2^2-c_3^2} }{c_3-c_1}
\end{equation}
\begin{equation}
\label{eq:phipm}
        \phi_{\pm}(v) = 2 \arctan\left(\frac{- c_2 \pm \sqrt{c_1^2+c_2^2-c_3^2} }{c_3-c_1}\right).
\end{equation}

We then used these solutions to find the region of $\phi$ over which to integrate. First, if Equation \ref{eq:phipm} had no solutions and if $c_1 \cos(\phi) + c_2 \sin(\phi) + c_3 > 0$ for any value of $\phi$, then the integration limits were $(\frac{\pi}{2},-\frac{\pi}{2})$. If $c_1 \cos(\phi) + c_2 \sin(\phi) + c_3 < 0$, then the limits were $(0,0)$. If there were solutions, we computed the integral over the region of $\phi$ between $\frac{\pi}{2}$ and $-\frac{\pi}{2}$ where $c_1 \cos(\phi) + c_2 \sin(\phi) + c_3>0$. This region was between or outside of $\phi_+(v)$ and $\phi_-(v)$. If the width of the integrated region was $\Delta\phi_{\text{illum}}(v)$, the un-normalised kernel was

\begin{equation}
\label{eq:cresent_kernel}
        k(v) = \Delta\phi_{\text{illum}}(v)\left(1 - \frac{v^2}{R^2(\wy^2 + \wz^2)}\right)^{\frac{1}{2}}.        
\end{equation}

\section{Mapping orbital parameters to the kernel}
\label{sec:toparmaters}

\subsection{Stellar rotational broadening}
\label{sec:map_stellar}

We modelled the star as a solidly rotating sphere with a uniformly bright disc so its rotational broadening kernel, $k_\star(v)$, is given by Equation \ref{eq:uniform_kernel}, with $R$ as the radius of the star, $R_\star$. We note that stars neither have uniformly bright discs nor rotate as solid bodies, but information on the deviations from these assumptions for stars other than the Sun is limited. Hence, we chose to make this assumption for now. For non-uniform brightness, we would need to evaluate Equation \ref{eq:unnormalised_kernel} for the chosen $I(\phi)$, while for non-solid body rotation, we would need to derive the kernel for spatially varying $\vec{\omega}(x,y,z)$. We discuss the impact of non-uniform brightness and non-solid body rotation in further detail in Section \ref{sec:approx_discussion}.

\begin{figure}
 \centering
  \includegraphics[width=.8\linewidth, trim={3cm 4cm 3cm 1cm},clip]{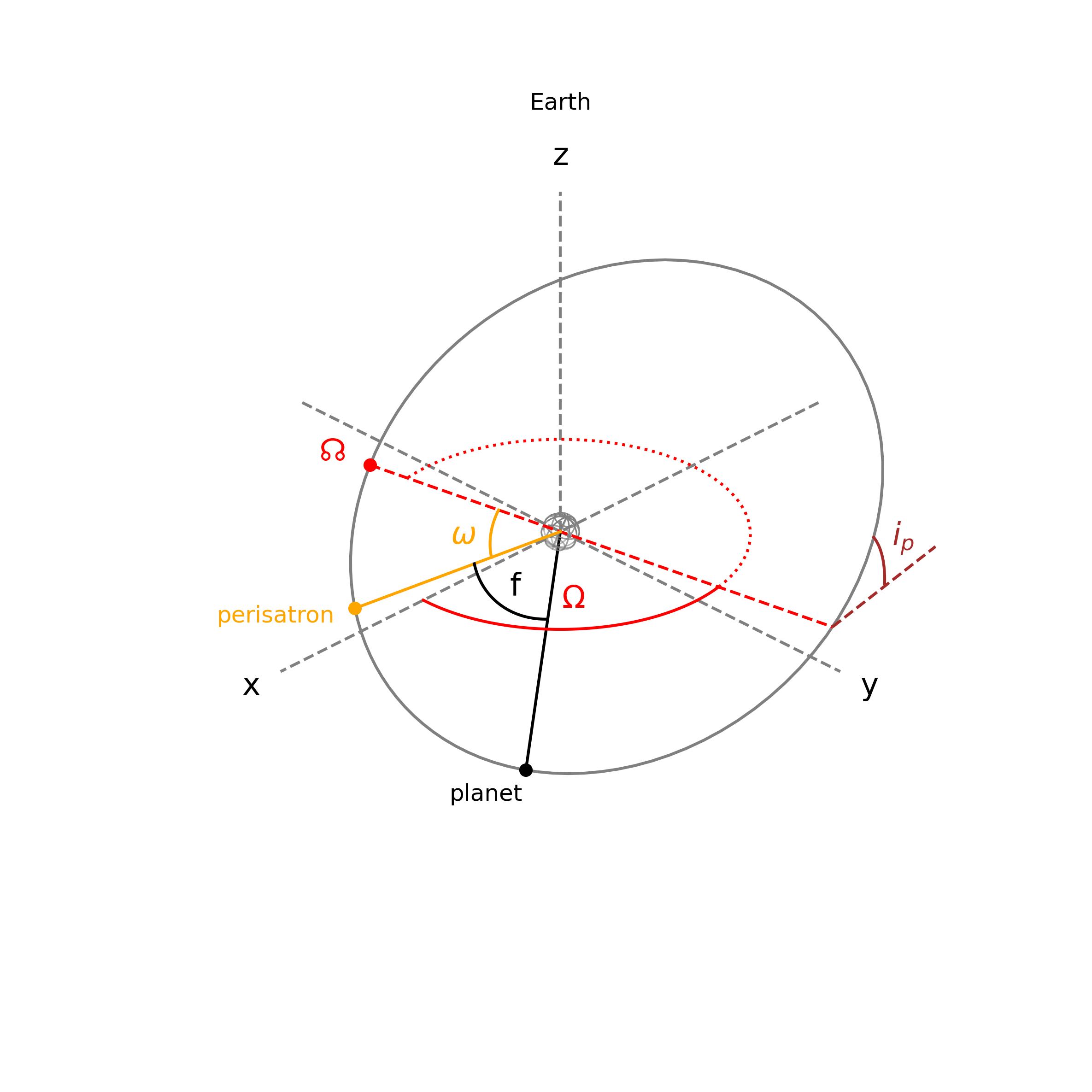}
    \caption{Angles defining the orientation of a planet's orbit.}
    \label{fig:planet_orbit}
\end{figure}

In Equation \ref{eq:uniform_kernel}, $\wy$ and $\wz$ are the $y$ and $z$ components of the angular velocity of the star, as seen by an instantaneously stationary observer on the positive $x$ axis. For the stellar rotational broadening, the observer is the planet. We describe the position of the planet as a function of time by

\begin{equation}
        \vec{r}_p = \frac{a(1-e^2)}{1+e\cos(f)}R_z(\Omega_p)R_x(-i_p)R_z(\omega_p^\circ) 
        \begin{bmatrix} \sin(f) \\ \cos(f) \\ 0 \end{bmatrix},
\end{equation}

where the orbital elements are $\Omega_p$, the longitude of the ascending node; $i_p$, the inclination; and $\omega_p^\circ$, the argument of periastron\footnote{We added a small circular symbol to the argument of periastron to clearly distinguish it as an angle rather than an angular velocity.}. Additionally, $f$ is the true anomaly, $a$ is the semi-major axis, and $e$ is the eccentricity. Figure \ref{fig:planet_orbit} shows the orbital elements. Here, $R_x(\alpha)$ and $R_z(\alpha)$ are

\begin{equation}
    R_x(\alpha) = \begin{bmatrix} 1 & 0 & 0 \\ 0 & \cos\alpha & -\sin\alpha \\ 0 & \sin\alpha & \cos\alpha \end{bmatrix} \quad R_z(\alpha) = \begin{bmatrix} \cos\alpha & -\sin\alpha & 0 \\ \sin\alpha & \cos\alpha & 0 \\ 0 & 0 &1 \end{bmatrix}.
\end{equation}

When all orbital elements are $0^\circ$, the planet is on the positive x axis, which is why we chose to place our observer there in the initial derivation.

\begin{figure}
 \centering
  \includegraphics[width=.8\linewidth, trim={3cm 4cm 3cm 1cm},clip]{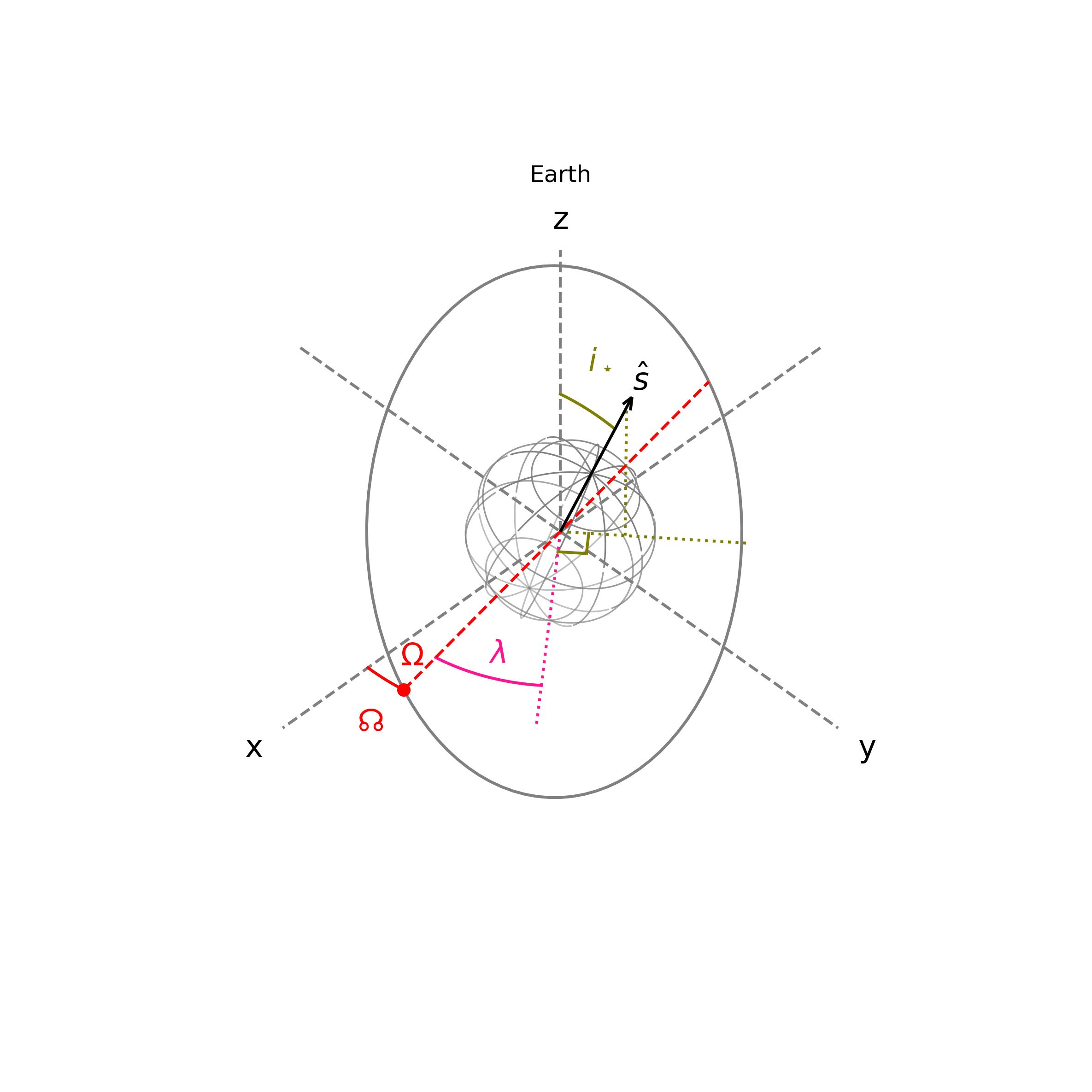}
    \caption{Angles defining the orientation of the stellar spin axis.}
    \label{fig:star_axis}
\end{figure}

In the same coordinate system, the stellar spin axis is defined by the stellar inclination $i_\star$ and the sky-projected obliquity $\lambda$ (shown in Figure \ref{fig:star_axis}) as follows:

\begin{equation}
    \vec{\omega}_\star = R_z(\Omega_p)R_z(\lambda)R_x(-i_\star) \begin{bmatrix} 0 \\ 0 \\ \omega_\star \end{bmatrix}.
\end{equation}

To transform into the frame in which the planet is instantaneously on the $x$ axis, we performed the following series of rotations: $R_z(-(\omega_p^\circ+f))R_x(i_p)R_z(-\Omega_p)$. We mark this frame with the subscript $p\rightarrow x$. The stellar angular velocity, $\vec{\omega}_{\star,p\rightarrow x}$, in this frame is 

\begin{equation}
    \vec{\omega}_{\star,p\rightarrow x} = R_z(-(\omega_p^\circ+f))R_x(i_p)R_z(-\Omega_p)R_z(\Omega_p)R_z(\lambda)R_x(-i_\star) \begin{bmatrix} 0 \\ 0 \\ \omega_\star \end{bmatrix}. 
\end{equation}

However, the planet has an instantaneous angular velocity $\vec{\omega}_{p}$ around the $z$ axis, given by

\begin{equation}
    \vec{\omega}_{p} = \sqrt{\frac{GM_\star(1+e\cos(f))^3}{a^3 (1-e^2)^3}} \begin{bmatrix} 0 \\ 0 \\ 1 \end{bmatrix}.
\end{equation}

We therefore transformed into the frame in which the planet is instantaneously stationary. We denote this frame by $p\rightarrow x,0$, where the zero indicates that the planet is stationary. Since angular velocity vectors are additive, the stellar angular velocity in this frame, $\vec{\omega}_{\star,p\rightarrow x,0}$, is

\begin{equation}
    \vec{\omega}_{\star,p\rightarrow x,0} = \vec{\omega}_{\star,p\rightarrow x} - \vec{\omega}_{p}.
\end{equation}

Evaluating this, we find that the $y$ and $z$ components of the star's angular velocity as viewed by the planet are

\begin{equation}
    \label{eq:wywz}
    \begin{split}
        \wy = & \omega_\star ( \sin(i_\star) \cos(f + \omega_p^\circ) \cos(i_p) \cos(\lambda) \\ &-\sin(i_p) \cos(f + \omega_p^\circ) \cos(i_\star) + \sin(i_\star) \sin(\lambda) \sin(f + \omega_p^\circ))\\
        \wz = & \omega_\star(\sin(i_p) \sin(i_\star) \cos(\lambda) + \cos(i_p) \cos(i_\star)) \\ &-\sqrt{\frac{GM_\star(1+e\cos(f))^3}{a^3 (1-e^2)^3}}.
    \end{split}
\end{equation}

We can substitute these into Equation \ref{eq:uniform_kernel} to obtain the stellar rotational broadening kernel.

\subsection{Planetary rotational broadening}
\label{sec:map_planet}

If we assume that the dayside portion of the planet's surface as viewed from Earth is uniformly illuminated and neglect the effect of atmospheric winds, then we can use the kernel in Equation \ref{eq:cresent_kernel}, with $R$ being the radius of the planet, $R_p$, to describe its rotational broadening, $k_p(v)$. Again, to account for non-uniform brightness, we would need to evaluate the integral for a different $I(\phi)$. To account for winds, we would need to derive the kernel starting from Equation \ref{eq:rotational_kernel} for spatially varying $\vec{\omega}(x,y,z)$. We discuss the impact of non-uniform brightness and non-solid body rotation in Section \ref{sec:approx_discussion}.

The kernel in Equation \ref{eq:cresent_kernel} requires the unit vector $\vec{\hat{n}}$, which points from the centre of the planet to the star, and $\vec{\omega}_{p,\circlearrowright}$, the angular velocity of the planet's rotation in a coordinate system where the observer is on the positive $x$ axis. However, in the conventional coordinate system, the observer, which for the planetary rotational broadening kernel is Earth, resides on the positive $z$ axis. Therefore, it is more convenient to calculate $\vec{\hat{n}}$ and $\vec{\omega}_{p,\circlearrowright}$ in a coordinate system where the observer is on the positive $z$ axis. As shown in Appendix \ref{sec:zobserver}, we re-derived the equations for an observer on the $z$ axis and obtained the same results, but with $n_x \rightarrow n_{z'}$, $n_y \rightarrow n_{y'}$, $n_z \rightarrow -n_{x'}$, $\wx \rightarrow \wzp$, $\wy \rightarrow \wyp$, and $\wz\rightarrow - \wxp$. Here, the primes indicate the observer is on the $z$ axis, while the unprimed coordinates correspond to an observer on the $x$ axis.

To calculate $\vec{\hat{n}}$ with the observer of the $z$ axis, we reversed and normalised the vector $\vec{r}_p$, which points from the centre of the star to the planet:

\begin{equation}
    \vec{\hat{n}} = - \vec{\hat{r}}_p = -R_z(\Omega_p)R_x(-i_p)R_z(\omega_p^\circ) \begin{bmatrix} \sin(f) \\ \cos(f) \\ 0 \end{bmatrix}
\end{equation}

\begin{equation}
\begin{split}
    n_{x'} &= \sin(\Omega_p)\sin(f+\omega_p^\circ)\cos(i_p)-\cos(\Omega_p)\cos(f+\omega_p^\circ)\\
    n_{y'} &= -\cos(\Omega_p)\sin(f+\omega_p^\circ)\cos(i_p)-\sin(\Omega_p)\cos(f+\omega_p^\circ)\\
    n_{z'} &= \sin(i_p)\sin(f+\omega_p^\circ).
\end{split}
\end{equation}

\begin{figure}
  \includegraphics[width=.8\linewidth, trim={4cm 5cm 5cm 4cm},clip]{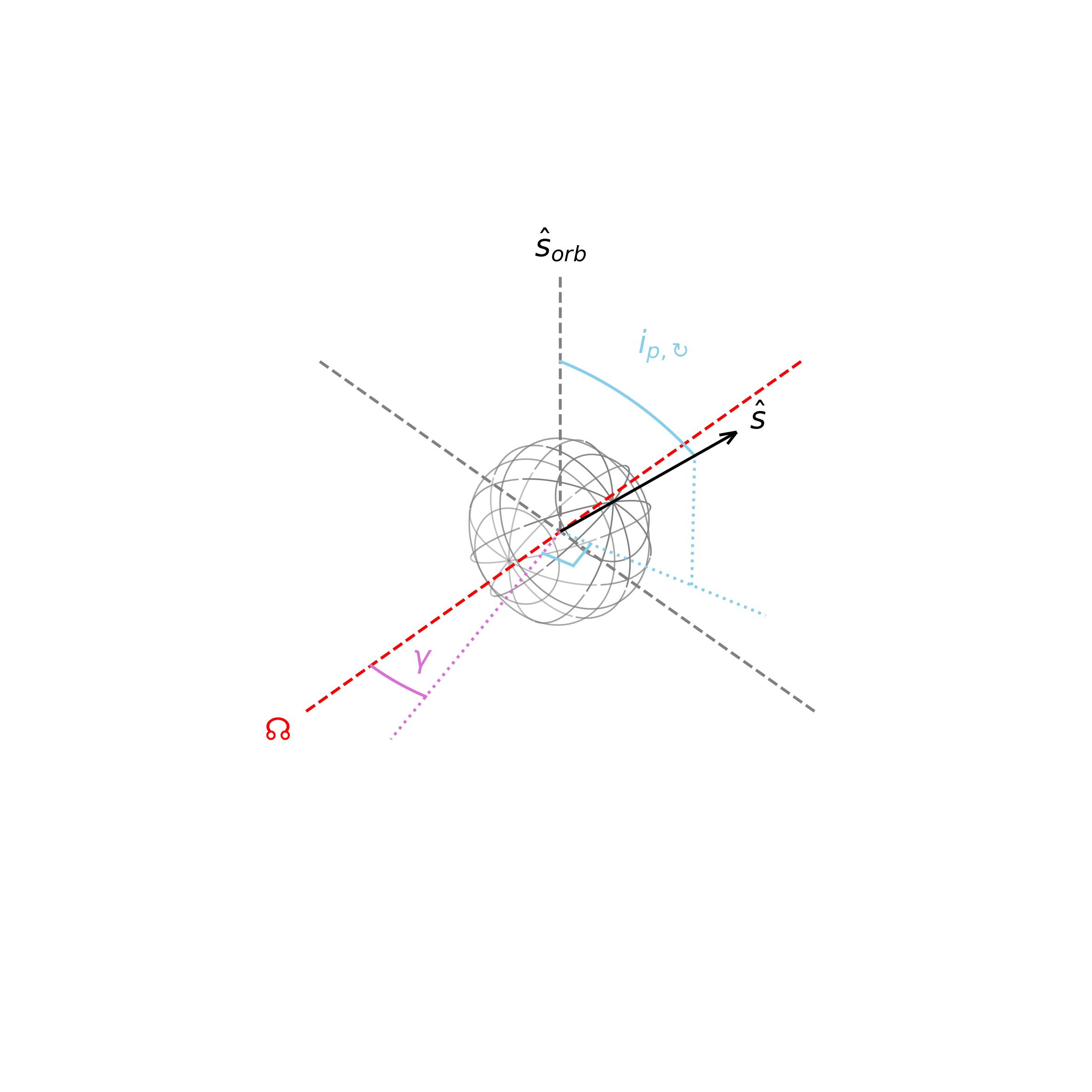}
    \caption{Angles defining the orientation of the planet's spin axis relative to the orbital spin axis.}
    \label{fig:planet_axis1}
\end{figure}

\begin{figure}
  \includegraphics[width=.8\linewidth, trim=4cm 5cm 5cm 3cm,clip]{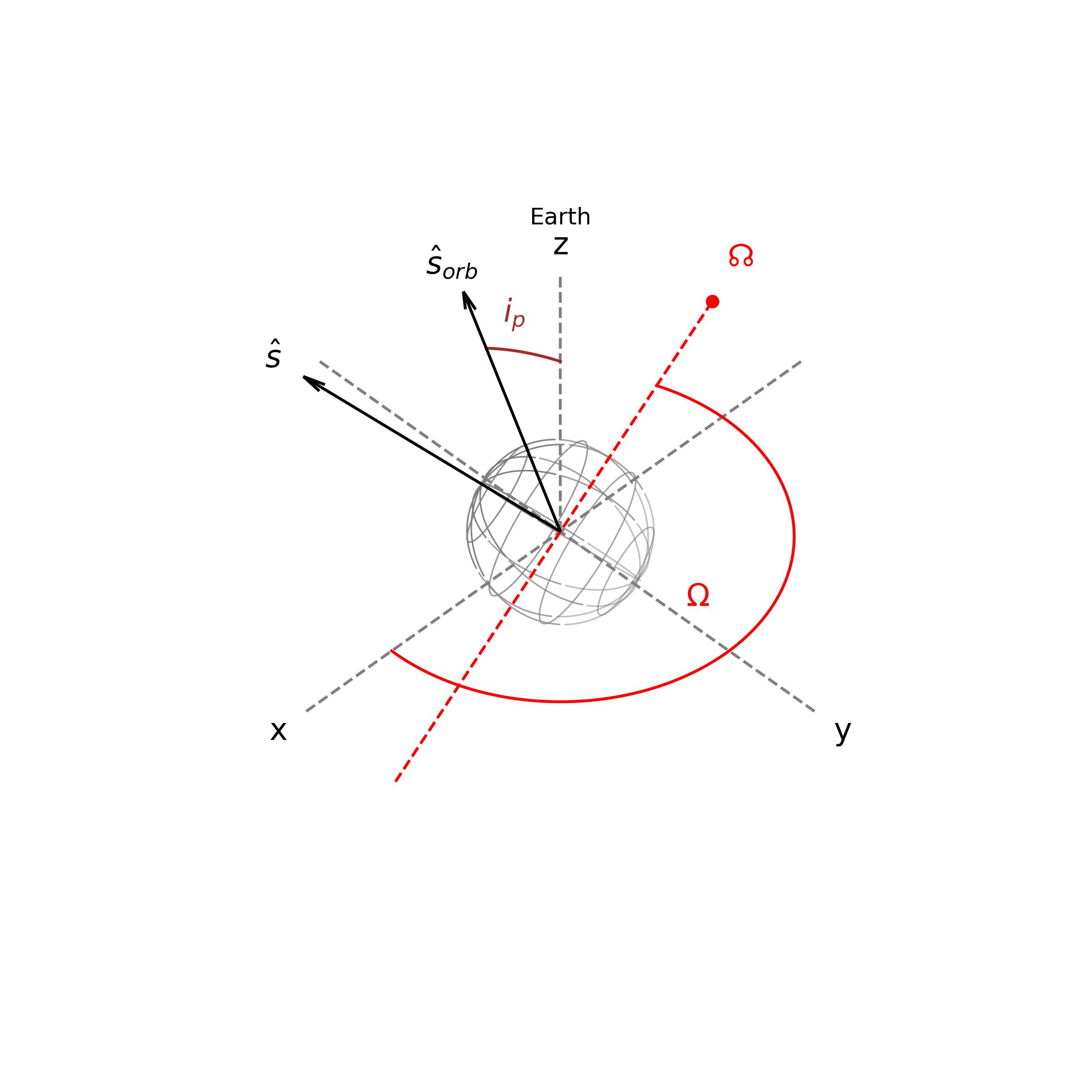}
    \caption{Angles defining the orientation of the orbital spin axis relative to the Earth.}
    \label{fig:planet_axis2}
\end{figure}

We substituted these into Equation \ref{eq:cresent_limit_equation_z} to find $\Delta\phi'_{\text{illum}}(v)$ in Equation \ref{eq:cresent_kernel_z}. To define the spin axis of the planet's rotation as seen from the Earth, we only need two angles. However, it is useful to define this axis relative to the orbital-spin axis so that it is easy to determine whether the spins were aligned. First, we defined a coordinate system in which the orbital-spin axis $\hat{s}_{orb}$ forms the $z$ axis, the line of nodes defines the $x$ axis, and the ascending node is on the positive $x$ axis. Figure \ref{fig:planet_axis1} shows the definition of the planet's spin axis by an anti-clockwise rotation of $\ips$ about the line of nodes, followed by a clockwise rotation of $\gamma$ about the orbital-spin axis. To rotate into the observer's frame, we first applied an anti-clockwise rotation $i_p$ around the $x$ axis (the line of nodes) and then a clockwise rotation of the line of nodes by an angle $\Omega$ around the $z$ axis. Figure \ref{fig:planet_axis2} shows this transformation. Therefore, the spin axis of the planet in the coordinate system with the observer on the positive $z$ axis is

\begin{equation}
       \vec{\omega}_{p,\circlearrowright} = R_z(\Omega_p)R_x(-i_p)R_z(\gamma)R_x(-\ips) \begin{bmatrix} 0 \\ 0 \\ \wps \end{bmatrix}
\end{equation}

\begin{equation}
\begin{split}
       \wxp =& -\wps(\sin(\ips)\cos(\gamma)\cos(i_p)\sin(\Omega) \\ &\qquad+ \sin(i_p)\cos(\ips)\sin(\Omega) + \sin(\gamma)\sin(\ips)\cos(\Omega) ) \\
       \wyp =& \wps(\sin(\ips)\cos(\gamma)\cos(i_p)\cos(\Omega) \\ &\qquad+ \sin(i_p)\cos(\ips)\cos(\Omega) - \sin(\gamma)\sin(\ips)\sin(\Omega))\\
       \wzp =& \wps(\cos(i_p)\cos(\ips) - \sin(i_p)\sin(\ips)\cos(\gamma)).
\end{split}
\end{equation}

We then substituted $\wxp$ and $\wyp$ into Equation \ref{eq:cresent_kernel_z} to obtain the planetary rotational broadening kernel.

\section{Discussion}
\label{sec:discussion}

\subsection{An example of these new kernels}
\label{sec:example}

In Sections \ref{sec:kernels} and \ref{sec:toparmaters}, we derived new rotational broadening kernels for the reflected-light spectra for a planet with arbitrary orbital and spin parameters. Here we show examples of how these vary with orbital phase. 

\begin{figure*}
    \centering
    \includegraphics[width=0.95\linewidth, trim={0cm 1cm 0cm 0cm},clip]{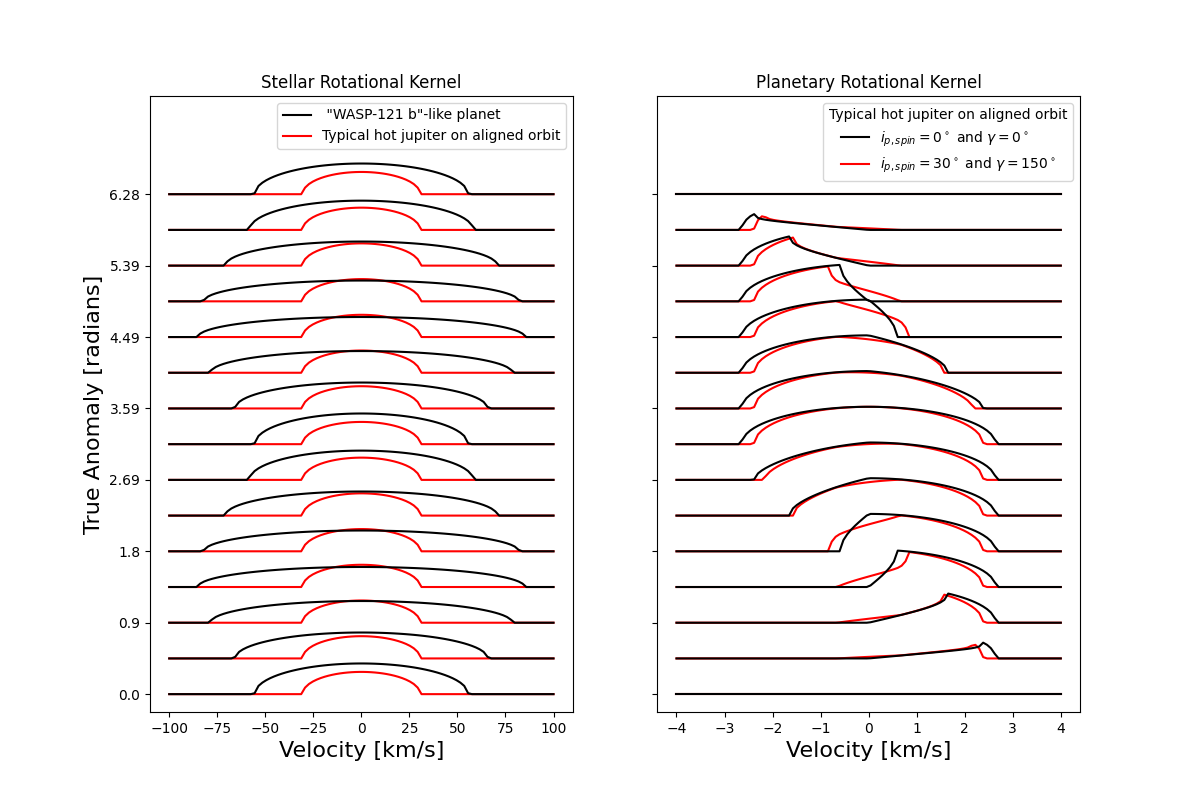}
    \caption{Example of the new rotational broadening kernels. Left panel: Stellar rotational broadening kernels for a WASP-121\,b-like planetary system and for a more typical hot Jupiter system with the planet on an aligned orbit. Right panel: Planetary rotational broadening kernel for a typical hot Jupiter with an aligned and non-aligned spin. Table \ref{tab:example_parameters} gives the full system parameters for the typical hot Jupiter and WASP-121\,b-like planetary system.}
    \label{fig:example}
\end{figure*}

For the stellar rotational broadening kernels, we show two models: one for a typical hot Jupiter on an aligned orbit and another largely based on the known parameters of the exoplanet WASP-121\,b \citep{Bourrier2020}. Table \ref{tab:example_parameters} summarises the parameters. WASP-121\,b provides an interesting test case because it is on a near-polar orbit around a rapidly rotating star. The left panel of Figure \ref{fig:example} plots the stellar kernels for the modelled planets as a function of the true anomaly. For the typical hot Jupiter, the kernel is narrower, with the rotational broadening velocity dominated by the planet's orbital motion. The kernel is also constant in time because the planet's view of the stellar rotation does not change throughout its orbit. For the WASP-121\,b-like planet, the kernel is much wider because the stellar rotation and planetary orbit are almost perpendicular to each other, meaning that their contributions to the rotational broadening velocity combine almost in quadrature. The kernel also changes with time, reaching its maximum width when the planet passes over the stellar equator and its minimum width when it passes over the poles. 

For the planetary rotational broadening kernels, we show two models of a typical hot Jupiter: one with an aligned spin and one with a spin that is misaligned with the orbit. The kernel for the non-aligned spin is slightly narrower because the line-of-sight component of the planet's rotational velocity is smaller. The shape of the kernel at crescent phases also changes because the velocities of the illuminated portion of the disc differ owing to the orientation of the spin axis. We note that the optical spectra of hot Jupiters are typically a mixture of reflection and thermal emission; therefore, the actual planetary broadening kernel may differ from that shown in Figure \ref{fig:example}. A non-uniform $I(\phi)$ could improve the accuracy of the modelled planetary rotational broadening kernel by accounting for the dayside hot spot produced by thermal emission. We discuss the impact of non-uniform $I(\phi)$ in Section \ref{sec:approx_discussion}.

\begin{table*}
\caption{Parameters assumed for the example planetary systems.}           
\label{tab:example_parameters}     
\centering                  
\begin{tabular}{l c c c c}   
\hline\hline            
Name & Typical hot Jupiter & WASP-121 & KELT-9 & Symbol \\  
\hline
Star & - & - \\
\hline
Radius, $R_\odot$ & 1.461 & 1.461 & 2.288 & $R_\star$ \\    
Rotation period, day & 28 & 1.13 & 0.667 & $\omega_\star = \frac{2 \pi}{P_\star}$ \\
Inclination of spin, deg & 90 & 8.1 & 52 & $i_\star$ \\
\hline
Orbit & - & - \\
\hline
Semi-major axis, au & 0.031 & 0.0257 & 0.0346 & $a$ \\
Eccentricity, - & 0 & 0 & 0 & $e$ \\
Period, day & 2.22 & 1.275 & 1.481 & $\omega_{p} = \frac{2 \pi}{P_p}$ $\dagger$ \\
Sky-projected Obliquity, deg & 0 & 87.2 & -84.8 & $\lambda$ \\
Inclination, deg & 90 & 87.6 & 86.79 & $i_p$ \\
Longitude of ascending node, deg & 90 & 90 & 90 & $\Omega_p$ \\
Argument of periastron, deg & -90 & -90 & 90 & $\omega_p^\circ$ \\
\hline
Planet & - & - \\
\hline
Radius. $R_{jup}$ & 1.13 & 1.742 & 1.891 & $R_p$ \\
Period, day & 2.22 & 1.275 & 1.481 & $\wps = \frac{2 \pi}{P_{p,\circlearrowright}}$ \\
Inclination of spin, deg & 0, 30 & 0 & 0 & $\ips$ \\
Spin obliquity, deg & 0, 150 & 0 & 0 & $\gamma$ \\
\hline                            
\end{tabular}
\newline $\dagger$ Used to calculate orbital angular velocity, as the orbit is circular.
\end{table*}

\subsection{Comparison with previous approaches}
\label{sec:comparison}

\subsubsection{Stellar rotational broadening}

Previous approaches to computing the stellar rotational broadening kernel have used the rotational broadening kernel for a uniformly bright sphere: 

\begin{equation}
    k_\star(v) \propto \left(1 - \frac{v^2}{v_{\text{rot}}^2} \right)^\frac{1}{2}, 
\end{equation}

where the proportionality indicates that a normalisation constant is required. Different approximations for the rotational broadening velocity, $v_{\text{rot}}$, have been used in this kernel. For a prograde aligned orbit, the rotational broadening velocity is constant throughout the planet's orbit and is given by \citep[e.g.][]{Rodler2010, Spring2022}

\begin{equation}
    v_{\text{rot}} =  2 \pi R_\star \left( \frac{1}{P_\star} - \frac{1}{P_p} \right),
\end{equation}

where the symbols have the same definitions as those in Table \ref{tab:example_parameters}. \citet{Winterhalder2026} observed the planet KELT\,9 b, which is on a near-polar orbit, at orbital phases between $0.58$ and $0.66$. To approximate the change in the rotational broadening velocity, they simplified the planet's orbit by assuming that it passed directly over the stellar poles at orbital phases $p=0.25$ and $0.75$ and crossed the stellar equator orbital phases $p=0$ and $0.5$. Since their $\omega_p^\circ=90^\circ$, the orbital phase is related to true anomaly via $f = 2 \pi p$. They then modelled the rotational broadening velocity using the projected stellar rotational broadening velocity given by

\begin{equation}
    v_{\text{rot}} = v_{\text{rot},\star} \left( \frac{1 - \cos(4 \pi p)}{2} \right),
\end{equation}

where $v_{\text{rot},\star} = \SI{111.4}{\kilo\meter\per\second}$ is the maximum rotational velocity measured by \citet{Gaudi2017}. This particular formulation neglects the contribution of the planet's orbital motion to the viewed stellar rotation and therefore to the rotational broadening. As a result, the rotational broadening reaches zero as the planet passes over the stellar pole.

In this work, we find the full analytical solution for the rotational broadening velocity is, where $\wy$ and $\wz$ are given in Equation \ref{eq:wywz},

\begin{equation}
    v_{\text{rot}} = R \sqrt{\wy^2 + \wz^2},
\end{equation}

\begin{figure}
  \includegraphics[width=.98\linewidth]{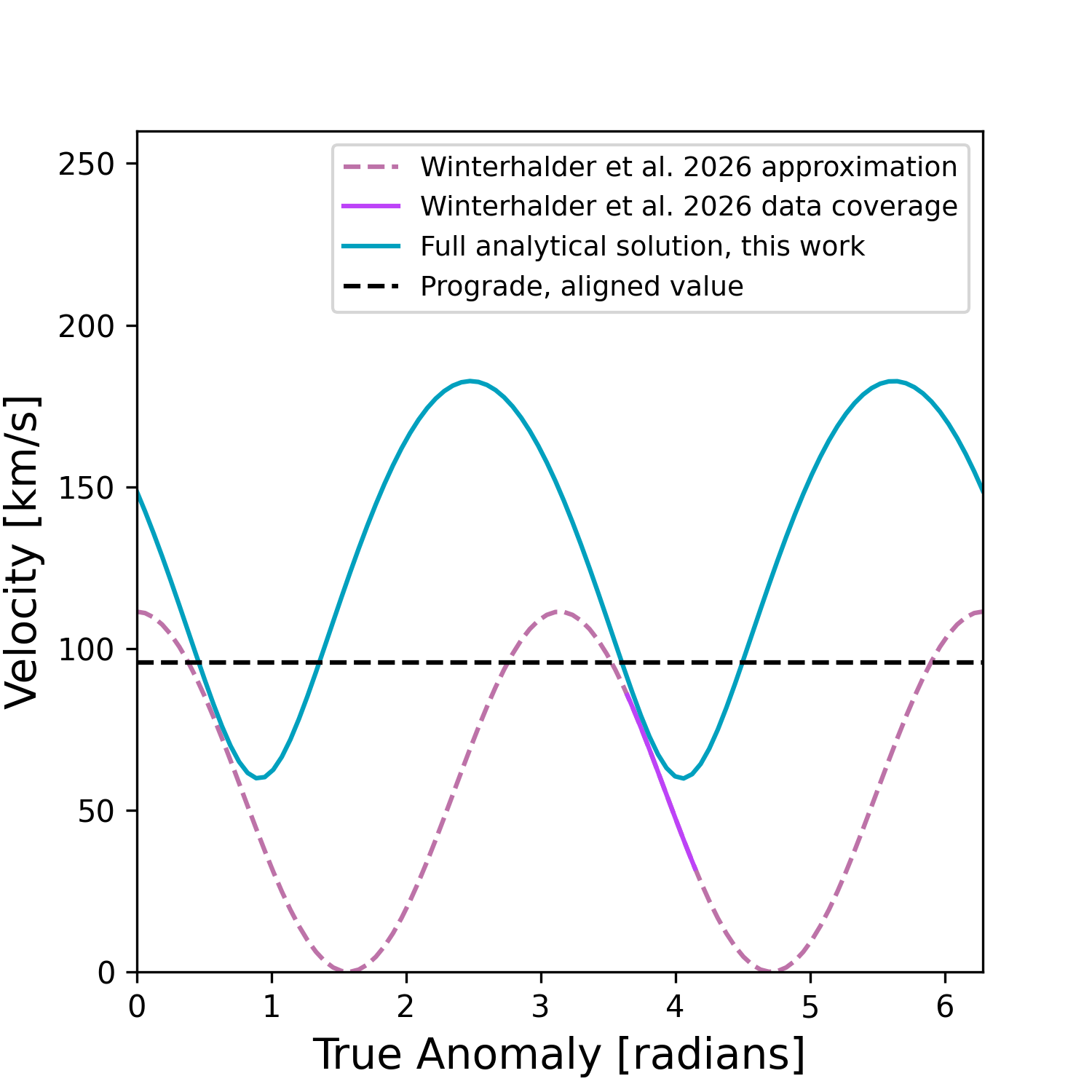}
    \caption{Comparison of previous approximations for the stellar rotational broadening velocity, $v_{\text{rot}}$ , with the analytical solution presented in this work for the reflected stellar light from KELT-9\,b.}
    \label{fig:kelt9}
\end{figure}

We used the KELT-9 system parameters presented in \citet{Gaudi2017} and \citet{Ahlers2020} (see Table \ref{tab:example_parameters}) to compute $v_{\text{rot}}$ for a prograde aligned orbit using the \citet{Winterhalder2026} approximation and the full analytical solution. Figure \ref{fig:kelt9} shows these results. Given the extreme parameters of this system, the rotational broadening velocity for the full analytical solution varies significantly over the orbit, with maxima as the planet transits the stellar equator and minima as it passes near the poles. The rotational broadening velocity never reaches zero because the planet's own orbital motion means that the star always appears to be rotating.

Comparison of the full analytical solution with that for the prograde aligned case shows that the orientation of a planet's orbit can strongly affect the stellar rotational broadening and that assuming the prograde aligned value can result in a significant under- or overestimation of the rotational broadening of the reflected stellar lines, depending on the observed orbital phase. The \citet{Winterhalder2026} approximation matches the full analytic solution for the first half of the orbital phases covered by their observations but then diverges. Their approximation does not provide a good prediction for the stellar rotational broadening velocity over the full orbit.

\subsubsection{Planetary rotational broadening}

\begin{figure}
  \includegraphics[width=.95\linewidth]{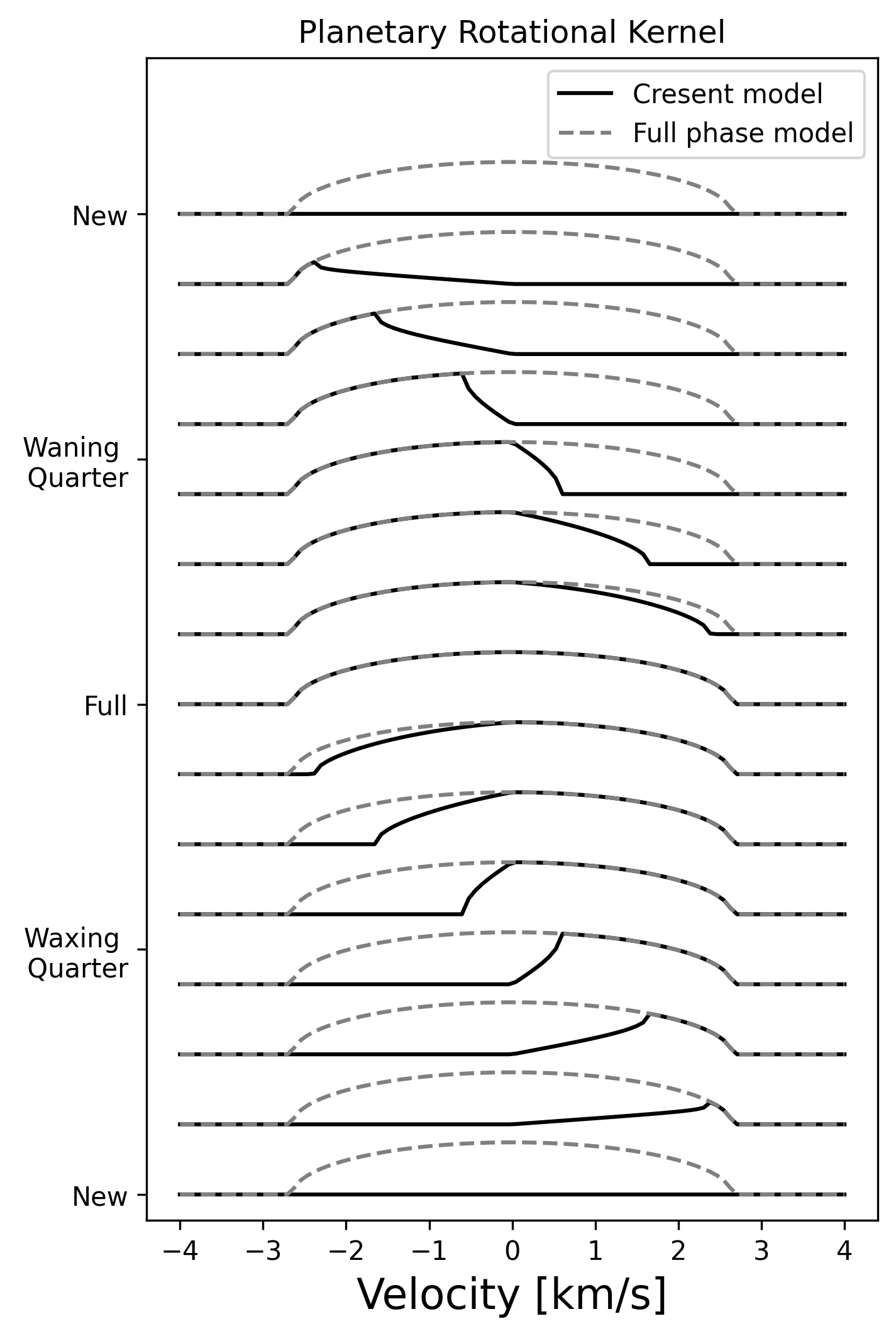}
    \caption{Comparison of the previous approximation for the full-phase planetary kernel with the analytical solution for a crescent presented in this work. These kernels correspond to the typical hot Jupiter with an aligned spin listed in Table \ref{tab:example_parameters}.}
    \label{fig:planet_comparison}
\end{figure}

For the planetary rotational broadening kernel, previous approaches have used the rotational broadening kernel for a uniformly bright sphere with a rotational broadening velocity that assumes the planet's spin is aligned with its orbit \citep[e.g.][]{Rodler2010, Spring2022} :

\begin{equation}
    k_p(v) \propto \left(1 - \frac{v^2}{v_{\text{rot},p}^2} \right)^\frac{1}{2} \quad \text{where} \quad v_{\text{rot},p} = 2 \pi \sin i_p \frac{R_p}{P_{p,\circlearrowright}},
\end{equation}

where the proportionality just indicates that a normalisation constant is required. In this work, we demonstrate in Figure \ref{fig:planet_comparison} how this kernel changes as the planet's dayside rotates in and out of view, the main effect of which is to narrow and offset the kernel. The narrowing of the kernel with phase is an important effect to consider when measuring the rotational period of an exoplanet from the rotational broadening of its spectral lines.

\subsection{Impact of approximations}
\label{sec:approx_discussion}

\subsubsection{Stellar limb darkening and differential rotation}

\begin{figure}
  \includegraphics[width=.85\linewidth]{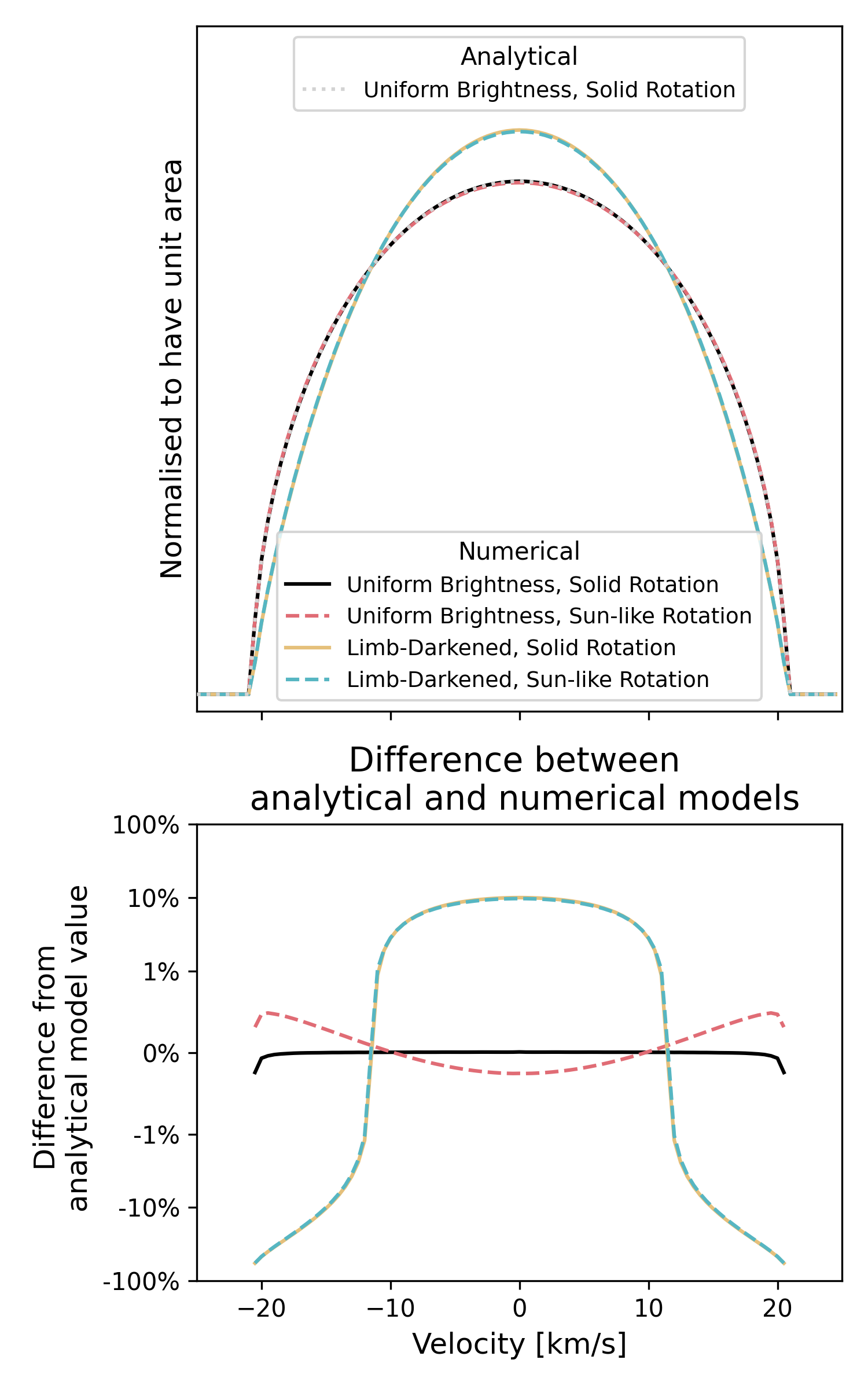}
    \caption{Numerically integrated stellar rotational broadening kernels containing differential rotation and limb-darkening, compared with the analytical solution for uniform brightness and solid-body rotation.}
    \label{fig:numerical_star}
\end{figure}

The analytical stellar rotational broadening kernels presented in this work assume that the stellar disc is uniformly bright and that the star rotates as a solid body. To investigate the impact of these assumptions, we numerically integrated Equation \ref{eq:rotational_kernel} for a typical hot Jupiter on a prograde aligned orbit (see Table \ref{tab:example_parameters}), accounting for these effects. We modelled the stellar limb darkening using the three-parameter non-linear law from \citet{Sing2010}, with the associated Convection, Rotation and planetary Transits (CoRot) stellar limb darkening coefficients for a $\SI{6000}{\kelvin}$ star. For differential rotation, we used the measurement of the solar angular velocity as a function of the stellar latitude presented in \citet{Snodgrass1990}. Figure \ref{fig:numerical_star} shows the numerically integrated kernels with each of these effects included, as well as both together, alongside the analytical model. The bottom panel shows the percentage difference between the value of the numerical kernel and the analytical solution. 

Differential rotation does not significantly change the kernel, with differences from the analytical solution remaining below $1\%$  This is because, for a hot Jupiter, the rotational broadening is dominated by the planet's orbital motion rather than the rotation of the stellar surface. For longer-period planets, which have smaller orbital-motion components, this limb darkening has a larger effect, but the kernel is also narrower, requiring extremely high spectral resolution (R$\sim$1,000,000) to resolve. Therefore, this effect is unlikely to have a strong impact on future observations. However, stellar limb darkening has a much stronger impact. Over most of the kernel, the difference from the analytical solution is less than $10\%$,  but the error increases significantly at the edges of the kernel. When limb-darkening is included, the kernel is narrower than the analytical solution. This does not have a significant impact on current high-resolution spectroscopy, as the kernel only needs to be approximately the right width to extract the signal of the reflected stellar lines in the data at low signal-to-noise. However, to infer the rotational properties of the system from the average stellar line shape, this effect must be considered. 

\subsubsection{Variation in planetary albedo}

\begin{figure}
  \includegraphics[width=.85\linewidth]{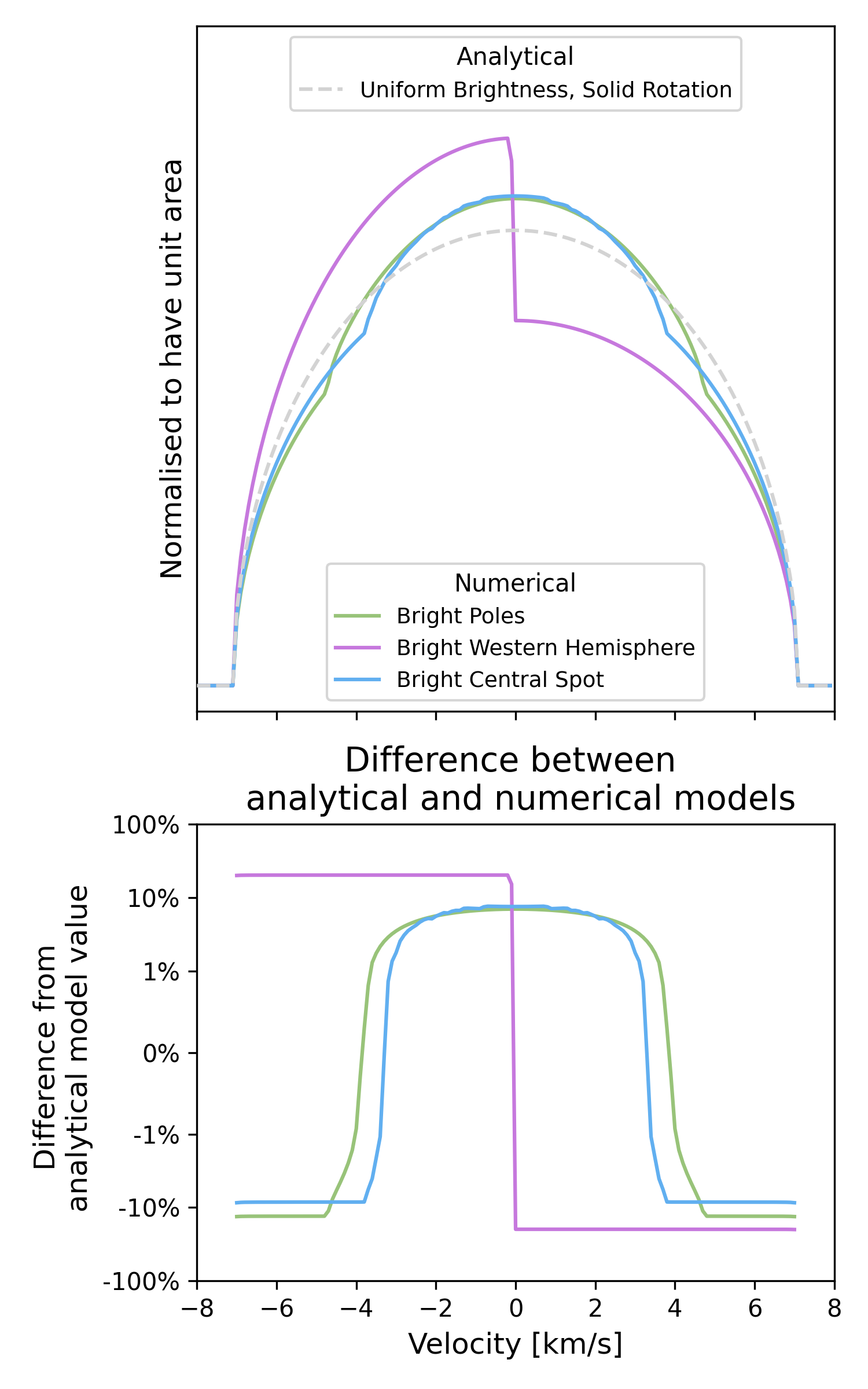}
    \caption{Numerically integrated planetary rotational broadening kernels for a solidly rotating planet with an aligned spin at full phase, with different albedo distributions across its visible surface, compared with the analytical solution for a uniformly bright disc.}
    \label{fig:numerical_planet}
\end{figure}

The analytical solution for the planetary rotational broadening kernel assumes that the planet has a uniform albedo across its visible disc. To investigate how a non-uniform albedo changes the kernel, we numerically integrated Equation \ref{eq:rotational_kernel} for a typical hot Jupiter, assuming that it is observed at full phase and that its spin axis is perpendicular to the line of sight. We created three models with different patches that are $1.5$ times the brightness of the rest of the disc. To model a planet with bright polar caps, we increased the brightness of regions obeying $|z| > 0.75R_p$; to model a planet with a brighter western hemisphere \citep[like LTT-9779\,b; ][]{Coulombe2025}, we selected regions with $y > 0$; and to model a bright central spot, we brightened regions satisfying $x > 0.85R_p$. Figure \ref{fig:numerical_planet} compares these numerical kernels with the uniformly bright analytical solution. For the polar-cap and bright-spot models, the kernel shape does not change dramatically, and the percentage difference remains within 10\% of the analytical model. The bright western-hemisphere model changes the kernel more significantly because the brightness variation aligns with the velocity field. Given the narrow width of these kernels, resolving these features would require extremely high spectral resolution (R$\sim$1,000,000)  and high signal-to-noise in the cross-correlated planetary spectrum to see these effects. Current instrumentation does not meet these requirements; however, these may need to be considered with future instrumentation. 

\subsubsection{Distance to the light source}

In this work, we ignored the finite size of the star when computing the stellar rotational broadening. \citet{Strachan2020} explored this effect in detail using numerical integration for tidally locked exoplanets on aligned circular orbits. In Fig. 15 of their work, the authors find that the finite size of the star has minimal impact for their hot Jupiter-type system for semi-major axes greater than approximately $\SI{0.05}{\astronomicalunit}$, or approximately nine times the stellar radius. For a semi-major axis around $4.5$ times the stellar radius, the finite size of the stellar disc results in an additional broadening of $\SI{10}{\kilo\meter\per\second}$. At smaller semi-major axes, the additional broadening increases rapidly. 

\begin{figure}
  \includegraphics[width=.98\linewidth]{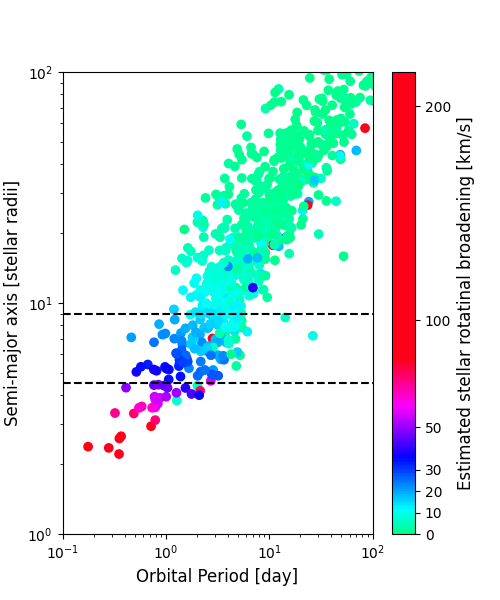}
    \caption{Predicted stellar rotational broadening velocity, assuming a prograde aligned orbit, for exoplanets in the NASA Exoplanet Archive \citep{Christiansen2025} around stars with measured rotation periods. The y axis shows the semi-major axis as a multiple of the host star's radius. Above the upper horizontal line, the finite size of the star has minimal effect on the broadening. The lower horizontal line indicates where an additional broadening of $\SI{10}{\kilo\meter\per\second}$ occurs.}
    \label{fig:nea}
\end{figure}

In Figure \ref{fig:nea}, we plot the predicted stellar rotational broadening velocity, assuming a prograde aligned orbit, for exoplanets in the NASA Exoplanet Archive \citep{Christiansen2025} that have a measurement of the stellar rotation period. At smaller semi-major axes, the estimated stellar rotational boarding velocity tends to increase. There are some exceptions, however, when the stellar rotation period is very short or similar to the planet's orbital period. The exoplanets around the line corresponding to nine times the stellar radius have a broadening of approximately $\SI{10}{}-\SI{20}{\kilo\meter\per\second}$ and those around the line corresponding to  $4.5$ times the stellar radius are broadened by approximately $\SI{50}{\kilo\meter\per\second}$. Therefore, in this region, the finite size of the star may typically contribute up to around $20$\% to the broadening for these close-in planets. However, at smaller separations, the rotational broadening due to the planet's motion increases rapidly, so the relative impact of the finite size diminishes. We highlight that for planet's orbiting within approximately nine times the stellar radius, this effect needs to be considered, but its relative contribution depends heavily on the system parameters. 

\section{Conclusions}
\label{sec:conclusions}

In this work, we derive the stellar and planetary rotational broadening kernels for exoplanets in reflected light. The former broadens the stellar spectrum incident on the planet based on the stellar rotation as  viewed by the planet, which is a combination of the star's rotation and the planet's orbital motion. The latter broadens the light reflected by the planet as a result of its spin. Sections \ref{sec:kernels} and \ref{sec:toparmaters} provide detailed derivations of these kernels, which assume solid-body rotation and no variation in surface brightness over the illuminated portion of the disc and exclude the finite size of the star for close-in exoplanets. In these sections, we highlight how our equations can be altered to account for the first two effects, and in Section \ref{sec:approx_discussion} we investigate their impact. Given the sensitivity of current instrumentation, these assumptions do not significantly hinder the study of exoplanets in reflected light. 

We also highlight that the equations presented in Sections \ref{sec:kernels} and \ref{sec:toparmaters} can also be used to compute other rotational broadening kernels. For example, we can obtain the kernel for an exoplanet's thermal emission spectrum from Equation \ref{eq:uniform_kernel} using an $I(\phi)$ that represents a hot spot. A numerical integration example of this is shown in Section \ref{sec:approx_discussion}. 

In Section \ref{sec:example}, we show examples of these kernels for an exoplanet on a near-polar orbit and for a typical hot Jupiter. These demonstrate that the shape of the broadening kernel can change significantly throughout the planet's orbit. In Section \ref{sec:comparison} we compare these results with previous approximations used in the literature, demonstrating the need for these analytical solutions. We highlight not only the need to account for this effect, but also that the rotational broadening of reflected-light spectra of exoplanets can be used to assess the alignment of systems, including those of non-transiting planets. This potentially exciting new avenue is worth considering when planning new observations with the ELT and beyond. 


\begin{acknowledgements}
      This research has made use of the NASA Exoplanet Archive, which is operated by the California Institute of Technology, under contract with the National Aeronautics and Space Administration under the Exoplanet Exploration Program.
\end{acknowledgements}

\bibliographystyle{aa}
\bibliography{aa61714-26.bib}

\begin{appendix}

\section{Proof that a convolution is equivalent to the kernel for finite lines}
\label{sec:proof_narrow_lines}

The un-normalized rotational kernel is defined as
\begin{equation}
    k(v) = \iint\limits_{A} I(x,y,z) \delta(v - v_{los}(x,y,z) ) \mathrm{d} A
\end{equation}
where $I(x,y,z)$ is the surface brightness of the sphere and $v_{los}(x,y,z)$ is the line-of-sight velocity. The integral is performed over a hemisphere facing the observer of radius $R$ which is centred at (0,0,0). Here, $\delta$ is a Dirac delta function, so the spectral lines are modelled as infinitely narrow. If the spectral lines have line shape $f(v)$ then the un-normalized line shape, including the rotational broadening, $g(v)$, would be

\begin{equation}
    g(v) = \iint\limits_{A} I(x,y,z) f(v - v_{los}(x,y,z)) \mathrm{d} A
\end{equation}

The convolution of $f(v)$ and $k(v)$ is,

\begin{equation}
    g(t) = \int\limits_{-\infty}^{\infty} f(\tau) \iint\limits_{A} I(x,y,z) \delta(t - \tau -v_{los}(x,y,z)) \mathrm{d} A \mathrm{d} \tau.
\end{equation}

Since $f(v)$, $I(x,y,z)$ and $\delta(\tau)$ are all positive real functions over the entire domain, the order of integration can be switched. Additionally, the integration limits do not need to be changed as the area integrated (that is the size of the star) is independent of the velocity. Therefore

\begin{equation}
    g(t) = \iint\limits_{A} I(x,y,z) \int\limits_{-\infty}^{\infty} f(\tau) \delta(t - \tau -v_{los}(x,y,z)) \mathrm{d} \tau \mathrm{d} A
\end{equation}

where we assume $I(x,y,z)$ is also independent of the velocity, which is equivalent to assuming that it does not change with wavelength. When integrated, Dirac deltas have the following property

\begin{equation}
    \label{eq:dirac_property}
    \int\limits_{-\infty}^{\infty} f(\chi)\delta(\chi - c) \mathrm{d}\chi = f(c)
\end{equation}

where $f$ is a function and $c$ is a constant with respect to $x$. This result applies as long as integration limits contain $x-c=0$. Therefore, the integral over $\tau$ is only non-zero when $t - \tau -v_{los}(x,y,z) = 0$. Thus, 

\begin{equation}
    g(t) = \iint\limits_{A} I(x,y,z) f(t - v_{los}(x,y,z)) \mathrm{d} A
\end{equation}

Thus, we see that $g(v) \equiv f(v) * k(v)$. Therefore, if we wanted to compute the line shape for non-infinitely narrow lines, we would simply need to convolve the chosen line-shape function with the rotational kernel $k(v)$.

\section{Re-derivation of kernels for an observer on the positive $z$ axis}
\label{sec:zobserver}

Here, we re-derive the kernels in Section \ref{sec:kernels} for an observer on the $z$ axis. For clarity, we will use prime notation for the coordinates to highlight the different location of the observer. We start from Equation \ref{eq:volume_integral} but this time with a line of sight velocity $v_{los}(x',y',z') = \wzp y' - \wyp x'$ and integrating over the observer facing hemisphere defined by $z'>0$.

\begin{equation}
\begin{split}
    k(v) = \int\limits_{0}^{\infty}\int\limits_{-\infty}^{\infty}\int\limits_{-\infty}^{\infty} I(x',y',z') \delta(v - \wxp y' + \wyp x') \\ \delta(x'^2+y'^2+z'^2 - R^2) \mathrm{d} x' \mathrm{d} y' \mathrm{d} z'
\end{split}
\end{equation}

We convert the integral into the following coordinate system

\begin{equation}
    \label{eq:z_coordinate_system}
    x = \frac{\wxp}{(\wyp^2 + \wxp^2)^\frac{1}{2}} r' \sin{\phi'} - \frac{\wyp v}{\wyp^2 + \wxp^2}; \quad y'=y'; \quad z' = r'\cos{\phi'}.
\end{equation}

Which results in the following integral (constants omitted)

\begin{equation}
    \begin{split}
        k(v) = \int\limits_{-\infty}^{\infty}\int\limits_{0}^{\infty}\int\limits_{-\frac{\pi}{2}}^{\frac{\pi}{2}} I(r', \phi', y') \delta(v - \wxp y' + \wyp x'(r',\phi')) \\ \delta(x'(r',\phi')^2+y'^2+z'(r',\phi')^2 - R^2)  r \mathrm{d}\phi'\mathrm{d}r'\mathrm{d}y'
    \end{split}
\end{equation}

As before, first we integrate over $y'$ which leads to

\begin{equation}
    \begin{split}
        k(v) = \int\limits_{0}^{\infty}\int\limits_{-\frac{\pi}{2}}^{\frac{\pi}{2}} I(r', \phi') \delta(r'^2 + \frac{v^2}{\wyp^2 + \wxp^2} - R^2)  r \mathrm{d}\phi'\mathrm{d}r'
    \end{split}
\end{equation}

We note here that this integral is only valid if  $\wxp \neq 0$. Integrating over $r'$ leads to (more constants omitted)

\begin{equation}
    \label{eq:unnormalised_kernel_z}
    k(v) = \int\limits_{-\frac{\pi}{2}}^{\frac{\pi}{2}} I(\phi')  \left( 1 - \frac{v^2}{R^2(\wyp^2+\wxp^2)} \right)^\frac{1}{2} \mathrm{d}\phi'.
\end{equation}

So the kernel is the same, but with $\wz^2$ replaced with $\wxp^2$ and the integral is now over $\phi'$. The relationship between $\phi'$ and $x'$, $y'$, $z'$ is slightly different to that between $\phi$ and  $x$, $y$, $z$. We apply the constraints $v - \wxp y' + \wyp x' = 0$ and $r'^2 + \frac{v^2}{\wyp^2 + \wxp^2} - R^2 = 0$, which are imposed during the evaluation of the integral, to the coordinate system in Equation \ref{eq:z_coordinate_system}.

\begin{equation}
    \label{eq:substitutions_z}
    \begin{split}
        x' &= \frac{\wxp}{(\wyp^2 + \wxp^2)^\frac{1}{2}} \left(R^2 - \frac{v^2}{\wyp^2 + \wxp^2}\right)^{\frac{1}{2}} \sin{\phi'} - \frac{\wyp v}{\wyp^2 + \wxp^2} \\
        y' &= \frac{\wyp}{(\wyp^2 + \wxp^2)^\frac{1}{2}} \left(R^2 - \frac{v^2}{\wyp^2 + \wxp^2}\right)^{\frac{1}{2}} \sin{\phi'} + \frac{\wxp v}{\wyp^2 + \wxp^2} \\
        z' &= \left(R^2 - \frac{v^2}{\wyp^2 + \wxp^2}\right)^{\frac{1}{2}} \cos{\phi'}.
    \end{split}
\end{equation}

Therefore the equivalent to Equation \ref{eq:cresent_limit_equation} is

\begin{equation}
\label{eq:cresent_limit_equation_z}
\begin{split}
    \frac{v(\wxp n_{y'} - \wyp n_{x'})}{\wyp^2 + \wxp^2} + n_{z'} \left(R^2 - \frac{v^2}{\wyp^2 + \wxp^2}\right)^{\frac{1}{2}} \cos(\phi') \\+ \frac{\wyp n_{y'} + \wxp n_{z'}}{(\wyp^2 + \wxp^2)^\frac{1}{2}} \left(R^2 - \frac{v^2}{\wyp^2 + \wxp^2}\right)^{\frac{1}{2}} \sin(\phi') = 0
\end{split}
\end{equation}

The solutions to this equation defines the integration limits for the rotational broadening kernel as before. So the kernel for a crescent in this new frame of reference becomes

\begin{equation}
\label{eq:cresent_kernel_z}
        k(v) = \Delta\phi'_{\text{illum}}(v)\left(1 - \frac{v^2}{R^2(\wyp^2 + \wzp^2)}\right)^{\frac{1}{2}}        
\end{equation}

Comparing Equations \ref{eq:unnormalised_kernel_z} and \ref{eq:cresent_limit_equation_z} with their counterparts in Equations \ref{eq:unnormalised_kernel} and \ref{eq:cresent_limit_equation}, we see they are identical with 

\begin{equation}
    \begin{bmatrix} n_{x'} \\ n_{y'} \\ n_{z'} \end{bmatrix} \rightarrow \begin{bmatrix} -n_{z} \\ n_{y} \\ n_{x} \end{bmatrix} \quad \text{ and } \quad \begin{bmatrix} \wxp \\ \wyp \\ \wzp \end{bmatrix} \rightarrow \begin{bmatrix} -\wz \\ \wy \\ \wx \end{bmatrix}
\end{equation}
    
\end{appendix}

\end{document}